\documentclass[prl,amsmath,superscriptaddress,showpacs,twocolumn]{revtex4}
\usepackage{graphicx}

\begin{document}

\title{Observation of Exciton-Phonon Sideband in Individual Metallic Single-Walled
Carbon Nanotubes}

\author{Hualing Zeng}
\affiliation{Department of Physics, The University of Hong Kong, Hong Kong, China}

\author{Hongbo Zhao}
\author{Fu-Chun Zhang}
\affiliation{Department of Physics, The University of Hong Kong, Hong Kong, China}
\affiliation{Center of Theoretical and Computational Physics, The University of Hong
Kong, Hong Kong, China}

\author{Xiaodong Cui}
\affiliation{Department of Physics, The University of Hong Kong, Hong Kong, China}

\date{\today}

\begin{abstract}
Single-walled carbon nanotubes (SWCNTs) are quasi-one-dimensional systems
with poor Coulomb screening and enhanced electron-phonon interaction, and are good candidates
for excitons and exciton-phonon couplings in metallic state.
Here we report back scattering reflection experiments on individual metallic SWCNTs.
An exciton-phonon sideband separated by 0.19 eV from the first optical transition
peak is observed in a metallic SWCNT of chiral index (13,10), which provides
clear evidences of excitons in metallic SWCNTs. A static dielectric constant of 10 is estimated
from the reflectance spectrum.
\end{abstract}


\maketitle

Single-walled carbon nanotubes (SWCNTs) have been the topic of extensive
research for their potentially broad applications \cite{Baughman02} and
fundamental scientific interest. SWCNTs can be either semiconducting or
metallic depending on their diameters and chiral angles.
One of the important and challenging issues is the possible excitons in
SWCNTs.
Excitons are pairs of electrons and holes bound by
the attractive Coulomb interaction,
and are fundamental to our understanding of optical properties of semiconductors.
After a series of theoretical and experimental works, a consensus has been reached that
excitons are also prominent factors in optical transitions
of semiconducting SWCNTs \cite{Ajiki93,Spataru04,Perebeinos04,Zhao04,WangF05}.

Although excitons are common in semiconductors and insulators, excitons are not expected
in bulk metals because of the metallic strong Coulomb screening, which prohibits the
bound state of electrons and holes in three dimensional (3D) systems.
SWCNTs are quasi-1D systems with poor Coulomb screening.
The relaxed bounding criteria
and the much weaker screening in low dimension are in favor of the existence of excitons
in metallic SWCNTs, as predicted in recent theoretical studies \cite{Deslippe07,WangZDmetallic}.
Therefore, SWCNTs provide an excellent opportunity to study the possible excitons in
metallic states. In metallic SWCNTs, except for the one pair of bands
that touch (or nearly touch) the Fermi surface, all pairs of bands are
gapped and may form exciton-continuum manifolds \cite{Zhao06} as in semiconducting SWCNTs.
Optical measurements addressing excitons in metallic SWCNTs are challenging and only very few
experiments have been reported.
Very recently, Wang \textit{et al}.\  reported
absorption spectrum of an armchair metallic SWCNT with chiral index of (21,21).
From the line shape of the optical transition,
they identified the signature of exciton formation \cite{WangF07}.
On the other hand, Berciaud \textit{et al}.\  reported
absorption spectrum using photothermal heterodyne
detection. They found no exciton-phonon sideband and
suggested weak excitonic effect in metallic SWCNTs \cite{Berciaud07}.
The lack of experiments on metallic SWCNTs is partially due to the fact that
photoluminescence spectroscopy, which is widely used in semiconducting SWCNTs, can not be employed
in metallic systems because of their low fluorescence yield.
Resonant Raman scattering is also popular, but difficult to obtain
a wide-range spectrum which is usually needed to obtain the excitonic information.
In contrast, Rayleigh scattering spectroscopy monitors optical properties
through scattering resonant enhancement when incident photon energy resonates
with optical transition, and was demonstrated as a powerful tool to
study electronic structure of both semiconducting and metallic SWCNTs \cite{YuZH01,Sfeir04}.

In the present Letter, we report the back scattering reflection spectrum of
suspended individual metallic SWCNTs, using a technique developed basing
on Rayleigh scattering spectroscopy \cite{Zeng08}.
An exciton-phonon sideband separated by 0.19 eV from the first optical transition
peak is observed in a metallic SWCNT of chiral index (13,10), which provides
clear evidence of excitons in metallic SWCNTs.
Our semi-empirical correlated calculation shows very good agreement of exciton
scenario with experiment, and estimates the static dielectric
constant of the SWCNT to be 10.

\begin{figure*}
\includegraphics[clip,width=5in]{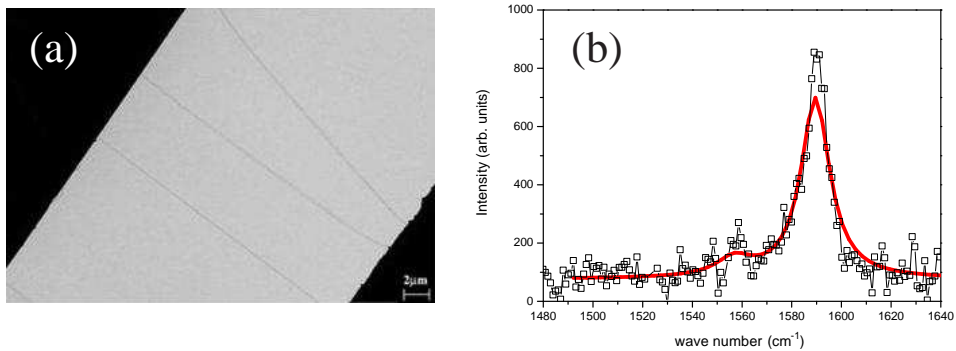}
\caption{\label{fig:Expt}
(Color online)
(a) SEM image of three isolated carbon nanotubes
suspended over the slit (white area) etched on silicon substrate (black areas
at two corners).
(b) G-mode resonant Raman spectrum of the (13,10) metallic SWCNT.
Open box symbols are experimental data and solid (red) line is a fit with
two Lorentzians.
}
\end{figure*}
The reflection experiment was performed on isolated individual SWCNTs suspended
over a single slit. A 20--30 $\mu$m wide 1 mm long slit on the silicon nitride
capped silicon substrate was fabricated with standard MEMS processes, including
silicon nitride etching mask growing by low pressure chemical vapor deposition
(LPCVD), optical lithography, reactive ion etching, and wet etching. Microscopically
long isolated SWCNTs were synthesized in situ across the slit by CVD method.
The catalyst was prepared by selectively dipping the diluted methanol solution
of FeCl$_{3}$ on the silicon substrate, followed by reduction in furnace
under an Ar/H$_{2}$ gas flow (400 sccm/50 sccm) at 900$^{\circ}$C for 20
minutes. Individual SWCNTs were formed on the substrate in the gas mixture
of Ar/H$_{2}$/ethanol at 900$^{\circ}$C for an hour. SWCNTs were aligned
perpendicular to the slit with the aid of the directional gas flow in the
CVD process, and some examples are shown in Fig.~\ref{fig:Expt}(a). A supercontinuum
light with wavelength from 500 nm to 2 $\mu$m was generated from an optical
crystal fiber pumped by a 100 mW pulsed Nd-Yag laser (pulse width $\approx1$
ns, repetition rate $=1$ kHz).
The incident light beam was tightly focused by a 40$\times$ achromatic
objective lens to a small spot of 4 $\mu$m on the SWCNTs, the back reflected
light off the SWCNTs was collected by the same lens. The reflectance spectrum
was obtained after corrections were made concerning incident light profile
and geometry dependent cross section. Details about the experimental setup
and data processing can be found elsewhere \cite{Zeng08}.

To identify the geometric structure of the SWCNTs,
we used combined technique of micro-Raman spectroscopy and Rayleigh scattering based
reflectance spectroscopy \cite{Heinz}.
Figure~\ref{fig:Expt}(b) shows the spectrum of the G-mode Raman scattering
of a SWCNT using a 632.8-nm excitation laser.
Full width at half
maximum (FWHM) of the peak is 8 cm$^{-1}$, indicating that this SWCNT is
isolated instead of among a small bundle \cite{Jorio02raman}. The asymmetric
line shape of the G-mode peak, which is the so called Breit-Wigner-Fano
line shape and can be better seen from the Lorentzian fitting shown in Fig.~\ref{fig:Expt}(b),
implies that
this SWCNT is metallic. The diameter of this SWCNT is estimated to be 1.58
nm, according to the empirical relation
$d=\sqrt{\frac{\omega_{G}^{+}-\omega_{G}^{-}}{C}}$,
where $\omega_{G}^{+}$ and $\omega_{G}^{-}$ are the frequencies of the
high- and low-energy G-mode sub-peaks in wave numbers, respectively, and $C=79.5$
cm$^{-1}$ is for metallic SWCNTs \cite{Jorio02Gband}.

\begin{figure}[b]
\includegraphics[clip]{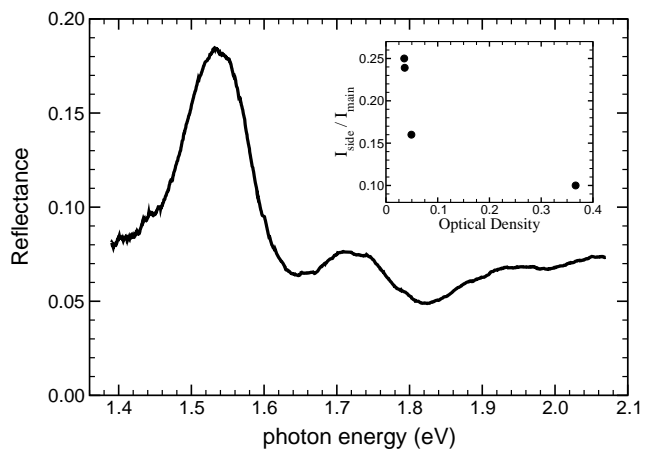}
\caption{\label{fig:Reflectance}
Experimental reflectance spectrum of the (13,10) metallic
SWCNT. Inset describes the intensity ratio of the sideband peak to the main
peak as a function of the incident light optical density (OD).
The average light intensity is $7\times10^{3}$ W/cm$^2$ at $\text{OD} = 0$.
}
\end{figure}
Figure~\ref{fig:Reflectance} shows the
Rayleigh scattering based reflectance spectrum of this SWCNT
in near-infrared to visible range \cite{Heinz}. There is a prominent peak at around
1.54 eV, followed by a weak sideband at $\sim$0.19 eV higher energy. The
first peak is from the first optical transition ($M_{11}$) for metallic SWCNTs.
Using Kataura plot \cite{Kataura99}, the only candidates having
the specified diameter and $M_{11}$ energy are the (13,10)
and (17,5) NTs. The latter option is, however, easily excluded on the
basis that its $M_{11}$ is splitted by more than 0.1
eV, due to trigonal warping effect \cite{Saito00}. Experimental FWHM
of the $M_{11}$ peak is only about 0.1 eV, which is consistent with the
(13,10) NT whose $M_{11}$ splitting is about 0.03 eV. Thus, we identified the
geometric structure of this metallic SWCNT to be (13,10).

Free electron band-to-band transition would give a broad and unstructured
reflectance spectrum of the $M_{11}$, and would feature
an asymmetric line shape with a long tail on the high energy side, reflecting
the $E^{-1/2}$ like density of states for quasi-1D systems.
The reflectance spectrum shown in Fig.~\ref{fig:Reflectance}, however,
presents a stark contrast. It has a well-defined \textit{single peak}, with
quite \textit{symmetric} line shape. Accordingly excitonic nature instead
of free band-to-band transition is speculated.

The strong evidence to support the exciton scenario lies in the broad sideband
located at 1.73 eV with a fraction of $M_{11}$ intensity.
The intensity ratio of the sideband peak to the main peak as a function of
the incident light optical density (OD) is nonlinear,
as shown in Fig.~\ref{fig:Reflectance} inset.
This effectively eliminates the possibility of the continuum origin of the sideband,
as the intensity ratio of continuum to an exciton would be independent of the
incident light OD.
The approximate energy difference of 190 meV between the sideband and $M_{11}$ is equal
to the energy of a zone-center G-mode phonon which is optically allowed.
Lacking any other physical processes
at this energy level, it is logical to attribute this sideband to the resonance
of the incident photon with the energy of the $M_{11}$
plus a G-mode phonon.
This picture could also explain Fig.~\ref{fig:Reflectance} inset.
We speculate that the nonlinear behavior is the result
of light heating effect which is due to extremely low heat capacity of the
suspended SWCNTs, and can strongly affect phonon population and exciton-phonon couplings \cite{thermal}.
Strong exciton-phonon coupling in SWCNTs has been proposed theoretically
\cite{Perebeinos05}, and has been shown experimentally being much
stronger than the coupling between free carrier and phonon \cite{Plentz05,Qiu05}.
The phonon sideband corroborates our conclusion that the main peak in reflectance
arises from exciton transition in the (13,10) metallic SWCNT.

Our result supports the argument in Ref.~\cite{WangF07}. The discrepancy
with Ref.~\cite{Berciaud07} may be explained by the different sample environment
in the measurements.
The samples here and in Ref.~\cite{WangF07} are suspended SWCNTs in air,
where the environmental dielectric constant is 1; the samples in Ref.~\cite{Berciaud07}
are SWCNTs isolated in surfactant micelles of SDS in D$_{2}$O, where the
dielectric constants are $\sim$1.5 and 80 for SDS and D$_{2}$O, respectively,
and a much stronger environmental screening is expected. We note that the
exciton-phonon sideband does not show up in the absorption spectrum in Ref.~\cite{WangF07}, which
could be due to weaker optical transition and shorter lifetime of the excitons associated with
the $M_{22}$ transition in their SWCNT.

To explain the experiment quantitatively, we have calculated reflectance spectrum
for (13,10) metallic SWCNT using semi-empirical $\pi$-electron model
which can capture the low-energy physics. Electron-electron (e-e) interactions are
included in the model, thus exciton physics is produced, and excellent agreements
with experiments have been achieved \cite{Zhao04,Zhao06,Wang06,WangZD07}.
The Hamiltonian is
\begin{equation}
 \begin{split}
   H= & -t\sum_{\left\langle ij\right\rangle \sigma} (c_{i\sigma}^{\dagger}c_{j\sigma}+\text{H.c.}) +U\sum_{i}n_{i\uparrow}n_{i\downarrow}\\
      & +\frac{1}{2}\sum_{i\neq j}V_{ij}(n_{i}-1)(n_{j}-1),
 \end{split}
\end{equation}
where $t$ is the nearest neighbor hopping, $U$ is the on-site Coulomb repulsion,
and $V_{ij}$ is the long range Coulomb interaction chosen to be in the Ohno
form \cite{Ohno64}
\begin{equation}
 V_{ij}=\frac{U}{\kappa\,\sqrt{1+0.6117\, R_{ij}^{2}}},
\end{equation}
where $R_{ij}$ is distance between atoms $i$ and $j$ in \AA, and parameter
$\kappa$ accounts for the dielectric screening. Taken into account that
SWCNTs are similar to $\pi$-conjugated polymers \cite{Zhao06} and that their
carbon atoms are non-planar, we chose $t=2.0$ eV \cite{Wang06}. We have
chosen $U=8$ eV, same as previous calculations on semiconducting SWCNTs
and $\pi$-conjugated polymers \cite{Chandross97b}.
In Ref.~\cite{WangZDmetallic}, $\kappa=3$ was selected for metallic SWCNTs, after fitting
calculated $M_{11}$ exciton energies to experimental ones for three metallic SWCNTs.
The same $\kappa=3$ was used in our calculation.
Configuration Interaction of
single excitations (SCI) from the Hartree-Fock (HF) ground state was used
to calculate the low-energy electronic structure and dipole coupling. For
(13,10) NT, we used open boundary condition and 4 unit cells,
which has more than 2000 carbon atoms. In SCI calculation, 100 HF levels
each from
above and below Fermi surface
are used, and convergence of the $M_{11}$ energy has been reached. We have also
tried Yukawa potential instead of the Ohno one, the result shows no significant
difference.

From the SCI calculated electronic structure, we can get the dielectric function
of an ideal macroscopic media consisting of aligned (13,10) metallic SWCNTs as \cite{HaugKochBook}
\begin{equation}
\epsilon(\omega)=\epsilon_{b}-\frac{4\pi n_{0}}{\hbar}
  \sum_j \left(\frac{|d_{j}|^{2}}{\omega-E_{j}+i\gamma}-\frac{|d_{j}|^{2}}{\omega+E_{j}+i\gamma}\right).
\end{equation}
Here, $E_{j}$ and $d_{j}$ are the energy and dipole coupling to the ground
state, respectively, of the $j$-th excited states obtained from the SCI calculation.
Background dielectric constant $\epsilon_{b}$, electron density of the media
$n_{0}$, and broadening $\gamma$ are the \textit{only three} parameters we used
to fit the experiment. $\epsilon_{b}$ should be unity if one has the complete
energy spectrum. Since only the low-energy spectrum are available from the
calculation, $\epsilon_{b}>1$ is used to account for the effect of the missing
higher energy oscillators. Reflectance can be calculated from dielectric
function using standard formula.

\begin{figure}
\includegraphics[clip,width=3.3in]{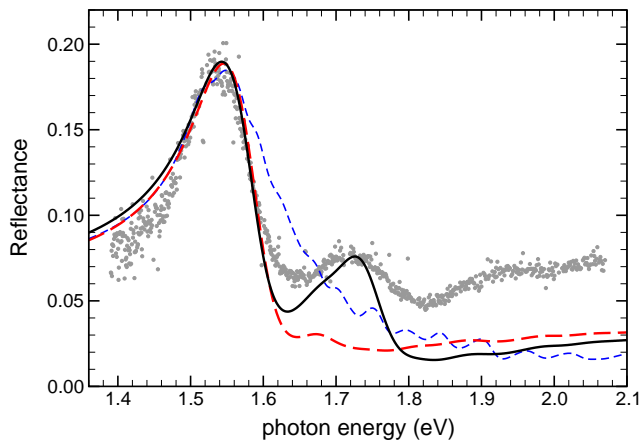}
\caption{\label{fig:theo}
(Color online)
Comparison of reflectance spectrum of the (13,10) metallic SWCNT between
experimental data (gray dots) and different calculations (lines). Short dashed
(blue) line is reflectance calculated from a tight binding model; long dashed
(red) line is from pure SCI calculation, and solid line is from the SCI calculation
with an added phonon 0.19 eV above the first exciton.
}
\end{figure}
In Fig.~\ref{fig:theo} the gray dots are the experimental reflectance data
of the (13,10) NT. Short dashed (blue) line is the reflectance
calculated from a tight binding model, with the nearest neighbor hopping
being 2.85 eV to match the calculated $M_{11}$ to the experiment. As discussed earlier,
we can clearly see here the asymmetric line shape featuring a long tail on the
high energy side. Tight binding calculation also produced much broader peak
than the experiment. The long dashed (red) line is the fit from our SCI calculation,
which has a very good agreement with the experiment for the $M_{11}$ peak.
This again shows that our model and calculation,
which include e-e interaction, can capture the low-energy physics
of SWCNTs. In the energy range shown in Fig.~\ref{fig:theo}, however, reflectance
from our SCI calculation shows only the $M_{11}$ peak. Our SCI calculation
does not include electron-phonon (e-ph) interaction, which means e-e
interaction alone can not explain the sideband, as we explain further below. Purely
electronic origin of the sideband could be the splitting of the $M_{11}$
exciton due to trigonal warping effect, or the continuum
band of the exciton. The SCI calculation showed that the splitting of $M_{11}$
is only about 20 meV, far less than the 190 meV separation. SCI calculation
did give the binding energy of $M_{11}$ exciton of (13,10) NT to be 0.2 eV,
which is consistent with the poor Coulomb screening in metallic SWCNTs,
and slightly smaller than the binding energy in semiconducting SWCNTs
with similar diameter.
However, as shown by the calculated spectrum, the continuum has very
weak oscillator strength, and can not produce a visible sideband.
The only other possible origin of the sideband is then the e-ph interaction
that was not included in our SCI calculation. With strong e-ph
interaction, it is possible that a phonon sideband shows up in the reflectance
spectrum. To test this idea, without changing any other parameters, we added
a single oscillator at the energy 0.19 eV above the $M_{11}$ exciton.
The solid line in Fig.~\ref{fig:theo} shows this fitting result. Both
the $M_{11}$ and the sideband can be fitted very well. Hence, our correlated
SCI calculation again showed the exciton nature of the main peak in reflectance
spectrum of (13,10) metallic SWCNT.

The fitting to the reflectance spectrum produced the dielectric function,
which provided an opportunity to estimate the static dielectric constant
$\epsilon_{0}$ of the metallic SWCNT. Dielectric constant is an important
material characterization and plays important role in manipulation and separation
of SWCNTs \cite{Krupke03}. Theoretical calculations
\cite{Benedict95,LinMF97screening,Leonard02}
and experiments have tried to establish the value, but the results vary in
large range. While a fitting of the charging energy gives $\epsilon_{0}$ as low as about
1.4 \cite{Tans97}, dielectrophoresis experiments suggest it could be as
high as few hundreds \cite{Krupke03}. There is also no optical experiment to determine
$\epsilon_{0}$. For (13,10) NT, the best fit to the reflectance spectrum
gives $\epsilon_{0}\approx10$. We note that due to difficulties in the
experiment, the absolute value of reflectance intrinsically carries a relatively
large error bar, which will affect the fitting result. Taking this into
account, we estimated that $\epsilon_{0}$ for (13,10) metallic SWCNT can be in range
between 7 to 15.

In summary, by performing a back scattering optical reflection experiment
on an isolated individual (13,10) metallic SWCNT suspended in air, we have observed a
single well-defined peak in the reflectance spectrum and a pronounced
exciton-phonon sideband
at a higher energy of 0.19 eV in the reflectance spectrum, which clearly shows
the excitonic nature of the optical transition in metallic SWCNTs.
Our theoretical calculations taking
into account of e-e interaction are in good agreement with
the excitonic picture, and give an estimation of the static dielectric constant of
$\epsilon_0 = 10$ for this SWCNT.

\begin{acknowledgments}
This work is supported by Hong Kong GRF grants HKU701907P and HKU706707P.
Computational resource from HKU's HPC Facilities is acknowledged.
\end{acknowledgments}

\end{document}